\documentclass[12pt]{article} 
\usepackage{authblk}
\usepackage{nrp} 

\begin{document}
\title{\textbf{\textsf{Superconductivity, Broken Gauge Symmetry, and the Higgs Mechanism}}}
\linespread{1.2}

\author{\textsf{Nicholas R. Poniatowski}\footnote{\texttt{nponiat@umd.edu}}}

\affil{\textit{\textsf{Department of Physics,}} \\ \textit{\textsf{Center for Nanophysics and Advanced Materials,}} \\ \textit{\textsf{University of Maryland, College Park, MD 20742}}}
\date{}

\maketitle 

\textsf{The association of broken symmetries with phase transitions is ubiquitous in condensed matter physics: crystals break translational symmetry, magnets break rotational symmetry, and superconductors break gauge symmetry. However, despite the frequency with which it is made, this last statement is a paradox. A gauge symmetry, in this case the U(1) gauge symmetry of electromagnetism, is a redundancy in our description of nature, so the notion of breaking such a ``symmetry'' is unphysical. Here, we will discuss how gauge symmetry breaks, and doesn't, inside a superconductor, and explore the fundamental relationship between gauge invariance and the striking phenomena observed in superconductors.}
\bigskip

\section{Introduction}
If one spends enough time running in condensed matter circles, one will surely hear the off-handed remark that superconductors break gauge symmetry. On the other hand, our undergraduate electromagnetism class taught us that a gauge symmetry is a mere redundancy, not a physical symmetry that can reasonably be ``broken'' in a consistent theory. So, what does it actually mean for a gauge symmetry to ``break,'' and why is this tied to superconductivity? In this article, we will address these questions, paying special attention to the role the mean field approximation plays in the study of superconductivity, and demonstrate how a classical field theory with spontaneously broken gauge symmetry can account for many of the spectacular properties of superconductors. 

The first four sections include preliminary information, enabling a reader with only a basic familiarity with quantum mechanics to follow the remainder of the discussion. We review what precisely electromagnetic gauge symmetry is in the in the second section, the notion of spontaneously broken symmetries in the third, and introduce second quantization and elements of quantum field theory in the context of superconductivity in the fourth section. The reader familiar with this material is encouraged to skip ahead to the main discussion of spontaneous gauge symmetry breaking inside superconductors which comprises the remainder of the article. Specifically, we will discuss the constraints gauge invariance imposes on any effective description of the superconducting state following Refs.  \cite{weinbergbook} and  \cite{weinbergpaper}, and finally carefully consider the Anderson-Higgs mechanism which is the crux of the gauge symmetry breaking story.

\section{Electromagnetic Gauge Symmetry}
Let us review how gauge symmetry arises in classical electromagnetism, \cite{jackson} starting with the Maxwell equations:
\begin{gather}
\nabla \cdot  \v{E} = \rho, \\ 
\nabla \cdot  \v{B} = 0 \label{max2}, \\
\nabla \times \v{E} + \frac{\del \v{B}}{\del t} = 0 \label{max3}, \\
\nabla \times \v{B} =\v{j} + \frac{\del \v{E}}{\del t}. \label{max4}
\end{gather}
Here and throughout the remainder of the paper we will use natural units where $\hbar = c = 1$, and further choose $\epsi_0 = \mu_0 =1$. The second and third equations may be satisfied by writing the physical $\v{E}$ and $\v{B}$ fields in terms of the electromagnetic potentials,
\begin{align}
\v{E} &= -\nabla \varphi - \deriv{\v{A}}{t}, \label{potdef1} \\
\v{B} &= \nabla \times \v{A}. \label{potdef2} 
\end{align}
However, the choice of $\varphi$ and $\v{A}$ is not unique for a given configuration of $\v{E}$ and $\v{B}$. Since $\nabla \times (\nabla \alpha) = 0$ for any scalar field $\alpha(\v{x},t)$, the transformation $\v{A} \mapsto \v{A} + \nabla \alpha$ leaves the magnetic field $\v{B} = \nabla \times \v{A}$ unchanged. To ensure the electric field is also invariant, we must simultaneously transform the scalar potential as $\varphi \mapsto \varphi - \del_t \alpha$. This intrinsic ambiguity in the electromagnetic potentials is called \emph{gauge symmetry}, and the transformation 
\begin{align}
\begin{split} \label{gaugetransf}
\v{A} &\mapsto \v{A} + \nabla \alpha, \\
\varphi &\mapsto \varphi - \deriv{\alpha}{t},
\end{split}
\end{align}
under which the electromagnetic field is invariant is called a \emph{gauge transformation}. Since our choice of the gauge parameter $\alpha$ is arbitrary, all physical observables must be independent of $\alpha$, that is, gauge invariant. 

To incorporate quantum mechanics, recall that in the presence of an electromagnetic field the momentum of a particle with charge $q$ must be shifted \cite{lagnote} $\v{p} \mapsto \v{p} - q\v{A}$ which corresponds to the replacing the operator $-i\hbar \nabla \mapsto -i\hbar \nabla - q\v{A}$ in quantum mechanics. The Schr\"odinger equation for such a charged particle in a potential $V(x)$ in the presence of an electromagnetic field is \cite{qmbook}
\begin{equation}
i\deriv{\psi}{t} = -\frac{1}{2m} (\nabla - iq\v{A})^2 \psi + (V + q\varphi)\psi. 
\end{equation}
If we gauge transform the fields while \emph{locally} rotating the phase of the wave function by the gauge parameter evaluated at each point in space and time,
\begin{equation}
\psi(\v{x},t) \mapsto \e^{iq \alpha(\v{x},t)}\psi(\v{x},t),
\end{equation}
the Schr\"odinger equation becomes \cite{derivnote}
\begin{align}
\begin{split}
i\left( \deriv{}{t} + iq\deriv{\alpha}{t} \right)\psi = &-\frac{1}{2m}\big(\nabla +iq\nabla \alpha - iq\v{A} - iq\nabla \alpha \big)^2 \psi \\
&+ \left(V + q\varphi - q\deriv{\alpha}{t} \right)\psi.
\end{split}
\end{align}
The $\alpha$ dependent terms drop out, and we recover the same Schr\"odinger equation we started with. Thus, the Schr\"odinger is gauge invariant under the quantum gauge transformation
\begin{align}
\begin{split} \label{qmgauge}
\psi(\v{x},t) &\mapsto \e^{iq\alpha(\v{x},t)} \psi(\v{x},t), \\
\v{A} &\mapsto \v{A} + \nabla \alpha, \\
\varphi &\mapsto \varphi - \deriv{\alpha}{t}.
\end{split}
\end{align}
Due to the phase rotation, this gauge symmetry is also said to be a local $U(1)$ symmetry. It is local because we may rotate the phase of the wave-function by a different amount at every point in space, in contrast to the weaker \emph{global} symmetry where we may only rotate the phase of the wave-function by the same fixed amount at every point. Further, $U(1)$ is the group of one-dimensional unitary matrices, which is equivalent to the set of all phases $\{\e^{i\theta}\}$ by which we may multiply the wave function. As in the classical case, all physically meaningful, observable quantities must be gauge invariant under the transformation in Eq. (\ref{qmgauge}) due to the arbitrariness of the gauge parameter.

\section{Spontaneous Symmetry Breaking}
Roughly speaking, a symmetry of a particular theory is a transformation which leaves the physics unchanged. More precisely, it is a transformation under which the Hamiltonian or Lagrangian defining the theory is invariant. We can explicitly break this symmetry by adding terms to the Hamiltonian or Lagrangian which do not respect the symmetry, corresponding to an external perturbation of the system. More interesting is the case of \emph{spontaneous symmetry breaking}, in which the system finds itself in a state that does not respect the symmetry of the underlying Hamiltonian or Lagrangian. \cite{ryder} For the purposes of this paper, we will say a symmetry is spontaneously broken when the ground state of a system is not invariant under a symmetry of the Hamiltonian or Lagrangian. 

This notion of symmetry breaking is central to the paradigmatic Landau theory of phase transitions. Since condensed matter systems are extraordinarily complex, diagonalizing the microscopic Hamiltonian for any interacting system is typically a hopeless endeavor. Instead, the Landau picture involves using the symmetries of the system to deduce its long distance behavior near a phase transition. In particular, we identify a symmetry respected by the phase on one side of the transition, and broken by the other (invariably, the ordered phase is less symmetric). Here, the symmetry we speak of is meant in the coarsest sense, since we have not introduced any Hamiltonian to appeal to. Instead, we quantify the symmetry by identifying an \emph{order parameter}: an observable whose expectation value vanishes in the symmetric phase and is nonzero in the symmetry breaking phase. 

To be concrete, consider a ferromagnet which, below the system's Curie temperature $T_c$, spontaneously magnetizes. Above $T_c$, the system is a paramagnet, with no net magnetization over any mesoscopic distance. In contrast, the ferromagnetic phase has a finite magnetization. Thus, the magnetization per unit volume $\v{m}(x)$ may serve as an order parameter. To identify the  symmetry broken at the transition, we note the system is isotropic in the paramagnetic phase, but the ferromagnet is magnetized along a particular direction in space. Thus, the ferromagnet spontaneously breaks rotational symmetry by ``choosing'' a magnetization axis. \cite{kardar}

Note however, that the magnetization axis is arbitrary. Thus, there is in fact a continuum of degenerate ground states: one corresponding to each possible magnetization direction. Furthermore, it costs no energy to uniformly rotate the system from one degenerate ground state into another. Then, by continuity, we expect that arbitrarily long wavelength variations in the magnetization direction will have an arbitrarily small energy cost. In the ferromagnet, these variations, called spin waves, dominate the low energy excitations of the system since they are gapless (their energy can be made vanishingly small). In fact, a general consequence of spontaneously broken \emph{continuous} symmetries is the emergence of gapless excitations around the ground state called \emph{Goldstone modes}, \cite{ryder} of which the spin wave is an example.

\section{Superconductivity} \label{scbackground}
A superconductor, as the reader is likely aware, is a phase of matter exhibiting a number of extreme phenomena, including the eponymous perfect conductivity and the expulsion of magnetic fields, or Meissner Effect. Microscopically, electrons are bound via an attractive interaction into Cooper pairs, which condense into a common macroscopic quantum state. 

Since superconductivity is an inherently quantum phenomenon as well as a many-body effect, it is most naturally described in the language of second quantization. We very briefly review this formalism in the following section, but the familiar reader is encouraged to skip this cursory treatment.
\subsection{Second Quantization}
Recall the harmonic oscillator from elementary quantum mechanics, the Hamiltonian of which may be ``factored" by introducing the operators,
\begin{align}
a = \frac{1}{\sqrt{2m\omega}}(m\omega \v{x} + i\v{p}),  \quad a^\dagger = \text{h.c.} 	
\end{align}
which are interpreted as raising and lowering the system between energy eigenstates $|n\rangle$, such that (neglecting normalizations), $a^\dagger |n\rangle \sim |n+1\rangle$ and $a|n\rangle \sim |n-1 \rangle$. Further, these eigenstates are orthonormal,
\begin{equation} \label{qhoortho}
\langle n | m \rangle = \delta_{nm}.
\end{equation}

Now, suppose we wish to consider a system that does not conserve particle number, \textit{i.e.} one in which particles may be created or destroyed. We can introduce analogous \emph{creation and annihilation operators}, 
\begin{align}
\Psi^\dagger (\v{x})& \quad \text{creates particle at point} \; \v{x}, \\
\Psi (\v{x})& \quad \text{annihilates particle at point} \; \v{x}.
\end{align}
In allowing for the particle number of our system to change, our space of possible states has expanded from a Hilbert space to a \emph{Fock space}, which can be thought of as collection of ``sectors" corresponding to a particular particle number, each of which is itself a Hilbert space. \cite{fock} Denoting an $N$-particle state as $|\alpha,N\rangle$ where the index $\alpha$ represents all other quantum numbers, we may write
\begin{align}
\Psi^\dagger |\alpha,N\rangle &= |\beta,N+1\rangle, \\ 
\Psi |\alpha,N\rangle &= |\beta, N-1 \rangle, \\
\langle \alpha, N|\beta,M\rangle &= \langle \alpha | \beta \rangle\; \delta_{NM}. \label{fockortho}
\end{align}
So we see the creation and annihilation operators take a state from one sector of the Fock space into another, and that states in different sectors of Fock space are orthogonal, \cite{Neignote} which is fairly intuitive since the particle number can only be changed by acting with the creation and annihilation operators. 

\subsection{Metals vs. Superconductors}
A metal, like most systems, has a definite particle number, so the expectation value of any operator which does not map a sector of the Fock space back onto itself will vanish. For example, choosing $\Psi_\sigma$ and $\Psi_\sigma^\dagger$ to create and annihilate electrons of spin $\sigma$,  
\begin{equation} \label{metalvev}
\langle \Omega_N | \Psi_\downarrow \Psi_\uparrow | \Omega_N \rangle = \langle \Omega_N | \Omega_N - 2\text{e}^- \rangle = 0,
\end{equation}
where $|\Omega_N\rangle$ is the ground state of the metal: perhaps a filled Fermi sea. Expectation values such as Eq. (\ref{metalvev}) are called \emph{ground state expectation values}, \emph{correlation functions}, or, in particle physics, \emph{vacuum expectation values} (VEV's) and are of central importance in field theories, as suggested by their many names (which we will use interchangeably). A common shorthand is to abbreviate these correlation functions as $\langle \Psi_\downarrow \Psi_\uparrow \rangle$ with the ground states implicit. Additionally, correlation functions such as $\langle \Psi_\downarrow \Psi_\uparrow \rangle$ in Eq. (\ref{metalvev}) which map a state to a different sector of Fock space are often called \emph{off-diagonal}, and as we saw vanish for any state of definite particle number. \cite{annett}

Now, let us a consider a superconductor, where a new ground state $|\Omega_S \rangle$ has developed. The pair condensate can be thought of as the superconducting vacuum (since paired electrons have minimal energy), which we denote as the field $\Delta(\v{x})$. The electrons in the superconductor may interact pairwise with the vacuum, either vanishing and creating a pair or appearing and breaking a pair, and as a result electron number is no longer conserved \cite{eqpnote} and states of different electron numbers may now have a nonzero overlap.

In particular, 
\begin{equation} \label{scvev}
\langle \Omega_S | \Psi_\downarrow \Psi_\uparrow | \Omega_S \rangle = \langle \Omega_S | \Omega_S - 2\text{e}^- \, +\text{pair}\rangle \neq 0.
\end{equation}
The ground state may now begin with $N$ electrons, interact with the condensate and exchange any number of pairs and electrons, and then be found in a state with $M$ electrons. Now, off-diagonal correlation functions such as Eq. (\ref{scvev}) no longer vanish and superconductors are said to possess \emph{off-diagonal long range order} (ODLRO). This electron number non-conservation is nowhere better illustrated than in the celebrated BCS wave function for the ground state of a superconductor, \cite{bcspaper}
\begin{equation}
|\Omega_{BCS}\rangle = \prod_{\v{k}} (v_{\v{k}}+ u_{\v{k}} c_{\v{k}\uparrow}^\dagger c_{-\v{k}\downarrow}^\dagger)|0\rangle,
\end{equation}
where $c_{\v{k}\sigma}^\dagger$ is the momentum space electron creation operator. Given the second term, we evidently have a state of indefinite particle number. 

Having seen the correlation function $\langle \Psi_\downarrow \Psi_\uparrow \rangle$ is zero in the normal phase and nonzero in the superconducting phase, it may serve as an order parameter for the superconducting state. \cite{colemangl} Following BCS, we define the superconducting order parameter,
\begin{equation} \label{scorderparam}
\Delta \propto \langle \Psi_\downarrow \Psi_\uparrow \rangle,
\end{equation}
 which we will later see coincides with the previously defined condensate field also labeled $\Delta$. 

Under a gauge transformation, creation and annihilation operators for charged particles must transform like the wave functions of charged particles in ordinary quantum mechanics. \cite{colemansc} For electron creation and annihilation operators (taking the charge of the electron to be $e = -|e|$), this means
\begin{align} \label{opgaugetrans}
\Psi_\sigma &\mapsto \e^{ie\alpha} \, \Psi_\sigma, \\
\Psi_\sigma^\dagger &\mapsto \e^{-ie\alpha} \, \Psi_\sigma^\dagger.
\end{align}
Consequently, operators which conserve electron number, such as the number operator $\Psi_\sigma^\dagger \Psi_\sigma$ are gauge invariant. Conversely, operators such as $\Psi_\downarrow \Psi_\uparrow$ which do not conserve electron number are \emph{not} gauge invariant, and transform as
\begin{equation} \label{psipsitrans}
\Psi_\downarrow \Psi_\uparrow \mapsto \e^{2ie\alpha}\; \Psi_\downarrow \Psi_\uparrow.	
\end{equation}
The expectation value, typically a real quantity, must then also transform as
\begin{equation} \label{vevtransf}
\langle \Psi_\downarrow \Psi_\uparrow \rangle \mapsto \langle \e^{2ie\alpha} \; \Psi_\downarrow \Psi_\uparrow \rangle = \e^{2ie\alpha}\langle \Psi_\downarrow \Psi_\uparrow \rangle.
\end{equation}
So, for our definition of the order parameter in Eq. (\ref{scorderparam}) to make sense, the order parameter itself must also transform as
\begin{equation} \label{orderparamgauge}
\Delta \mapsto \e^{2ie\alpha} \, \Delta.
\end{equation}
Thus, the order parameter is complex, and, more importantly, non-gauge invariant. Since all observable quantities must be gauge invariant, this implies the order parameter is \emph{not} observable (The magnitude of the order parameter $|\Delta|$, the energy gap in the superconductor's spectrum, is observable, however its phase is not). 

Note that the expectation value $\langle \Psi \Psi \rangle$ is non-gauge invariant in any system, not just a superconductor. However, such an expectation value is identically zero in an ordinary system, and zero transforms trivially under a gauge transformation. What makes a superconductor special is that this expectation value is nonzero, allowing its transformation properties to be physically relevant. 

\section{The Role of Mean Field Theory}
In the previous section we have seen that the superconducting order parameter is a gauge covariant quantity, which is to say that it transforms in a simple -- but nontrivial-- manner. In subsequent sections we will consider descriptions of superconductors which take $\Delta$ as the primary dynamical quantity, and its gauge covariance will be of central importance. As such, it is worth briefly discussing how it is possible for $\Delta$ to attain a nonzero expectation value, which is the origin all of the physics to follow. 

The microscopic theory of superconductivity is written in the language of quantum field theory, and the methods necessary for a quantitative treatment of this issue lie well outside the scope of this article. Here, we will sketch how such an analysis proceeds, and refer the reader seeking further detail to the literature. \cite{condmatfieldth,colemansc} 

A quantum field theory can be constructed from a quantum action, which (roughly speaking) specifies the energy associated with each allowed microscopic process. \cite{shankar} In a superconductor, one such process is the pairing interaction, which contributes a term
\begin{equation} \label{qmaction}
S_{\text{pair}} = g \,\Psi_\uparrow^\dagger \Psi_\downarrow^\dagger \Psi_\downarrow \Psi_\uparrow,
\end{equation}
where $\Psi_\sigma$ and $\Psi_\sigma^\dagger$ are the electron creation and annihilation operators introduced in section \ref{scbackground} and $g$ is the electron-phonon coupling constant. \cite{condmatfieldth} Unfortunately, such a quartic term in the action typically makes the quantum theory not analytically solvable. So, to progress we must approximate. In particular, we will make a mean field approximation, which capitalizes on the macroscopic, semi-classical nature of a superconductor. \cite{bruus}

To begin, we write a pair of electron annihilation operators as (in what follows we will omit spin indices for brevity)
\begin{equation} \label{mftdef}
\Psi \Psi = \langle \Psi \Psi \rangle + \Psi \Psi - \langle \Psi \Psi \rangle \equiv \langle \Psi \Psi \rangle + \eta.
\end{equation}
In the second equality, we interpret the operator $\Psi \Psi$ as being the expectation value $\langle \Psi \Psi \rangle$ plus quantum fluctuations, $\eta \equiv \Psi \Psi - \langle \Psi \Psi \rangle$ around it. Note, crucially, that $\Psi \Psi$ is an \emph{operator} which annihilates two electrons, whereas its expectation value is simply a \emph{number}. 

Substituting Eq. (\ref{mftdef}) and a similar expression for $\Psi^\dagger \Psi^\dagger$ into Eq. (\ref{qmaction}) gives
\begin{equation}
S_{\text{pair}} = g\langle \Psi^\dag \Psi^\dag \rangle \langle \Psi \Psi \rangle - g \langle \Psi^\dag \Psi^\dag \rangle \Psi \Psi - g \Psi^\dag \Psi^\dag \langle \Psi \Psi \rangle + \mathcal{O}(\eta^\dag \eta).
\end{equation}
For a macroscopic system such a superconductor, we may assume quantum fluctuations are in some sense small, and neglect terms of order $\eta^\dag \eta$. Defining the order parameter $\Delta \equiv g \langle \Psi \Psi \rangle$, we are left with
\begin{equation} \label{mfts}
S_{\text{pair}} = \frac{\Delta^\star \Delta}{g} - \Delta^\star \Psi \Psi - \Psi^\dag \Psi^\dag \Delta  .
\end{equation}
Since the order parameter field $\Delta$ is a number -- not an operator -- the second two terms in Eq. (\ref{mfts}) do not conserve electron number. This electron non-conservation allows the order parameter $\langle \Psi \Psi \rangle$ to acquire a nonzero vacuum expectation value, since the mean field action now connects sectors of the Fock space with different particle number via the last two terms. 

Thus, it is ultimately the mean field approximation which is responsible for electron number non-conservation, and consequently $\Delta$ acquiring a nonzero vacuum expectation value. However, notice that the action is still perfectly gauge invariant, given the transformations from Eq. (\ref{psipsitrans}) and (\ref{orderparamgauge}). Although the mean field approximation is implicit in both the BCS and Ginzburg-Landau theories of superconductivity that form the cornerstones of the discipline, it is worth noting that non-mean field and manifestly gauge invariant approaches to superconductivity exist such as the perturbative formalism developed by Nambu, \cite{nambu} as well as electron number conserving theories such as number projected BCS. \cite{nbcs1, nbcs2}

\section{Ginzburg-Landau Analysis} \label{glsection}
Having seen how it is possible for the order parameter $\Delta$ to acquire a nonzero expectation value in the superconducting state, let us consider the textbook treatment of symmetry breaking in superconductors: the Ginzburg-Landau theory, \cite{gloriginal} in which the free energy of the system is expanded as a series in $\Delta$ near the phase transition where $\Delta$ is small, such that we need only to keep the first few terms, \cite{saddleptnote}
\begin{align}
\begin{split} \label{glexp}
F = \int \df^3 x \; \bigg[ &\frac{1}{2m^\star}(\nabla + 2ie\v{A})\Delta^\star \cdot (\nabla - 2ie\v{A})\Delta \\
  &+ r \Delta^\star \Delta + \frac{u}{2} (\Delta^\star \Delta)^2 + \frac{1}{2}\v{B}^2 + \dots \bigg].
\end{split}
\end{align}
Here, $r$ and $u$ are unknown temperature dependent expansion coefficients, and the coefficient of the first term is conventionally chosen to be $1/2m^\star$, where $m^\star = 2m_e$ is the mass of a Cooper pair. To ensure the free energy is gauge invariant (and thus observable), the expansion includes only powers of the gauge invariant quantity $\Delta^\star \Delta$,  and in the ``kinetic'' term we have again replaced $\nabla \mapsto \nabla -2ie\v{A}$ (since the charge of a Cooper pair is $2e$). Finally, the $\v{B}^2$ term accounts for the energy of an external magnetic field.

The ground state may then be identified via a variational principle: it will be the state for which the free energy is minimized with respect to the order parameter. Since spatial variations of the order parameter are energetically costly, we may assume $\Delta$ is homogenous in the ground state of the system, and thus the first term in Eq. (\ref{glexp}) vanishes. Carrying out the variation imposes the condition 

\begin{equation}
0 = \fderiv{F}{\Delta^\star} = \big[r + u (\Delta^\star \Delta) \big]\Delta,
\end{equation}
which has two solutions: $\Delta = 0$, which we associate with the metallic phase above the phase transition, or $\Delta^\star \Delta = -r/u$, which we associate with the superconducting phase. \cite{rnote} Since this fixes only the amplitude of $\Delta$, we appear to have a continuum of ground states, parameterized by the phase angle $\phi$,
\begin{equation}
\Delta = \sqrt{\frac{|r|}{u}} \; \e^{i\phi}.
\end{equation}
At first glance, we appear to have a degenerate manifold of non-gauge invariant ground states, and are tempted to conclude that gauge symmetry has been spontaneously broken. However, as we will see, this apparent breaking of gauge symmetry is merely an illusion, whereby gauge redundancy and the covariance of the order parameter have conspired to make a single physical ground state appear to be an infinite set of distinct states. In fact, it is this very deception that is the origin of the misleading term ``broken gauge symmetry.''
After getting better acquainted with the subtlties of gauge invariance in the superconducting state in the next section, we will return to this state of affairs in section \ref{higgs}, and discover what it really means to ``break'' a gauge symmetry.

\section{Effective Field Theories} \label{eft}
In general, an effective field theory such as the Ginzburg-Landau theory discussed above involves trading the microscopic Hamiltonian of a system for an effective description that captures the low energy physics using emergent degrees of freedom, \cite{kardar} in this case the order parameter field $\Delta(\v{x},t)$, which we have now generalized to be time dependent. To construct an effective field theory we write down a Lagrangian, Hamiltonian, Free Energy, etc. based on symmetry arguments, and (classically) derive the resulting physics from the appropriate variational procedure. This central quantity, which we will call $\mathcal{F}$ and refer to as the ``effective action'' is generally a functional of the order parameter and its derivatives, and \textit{must} be gauge invariant (this is one of the symmetries we must take into account). In this section, we will show that the constraint that $\mathcal{F}$ be gauge invariant coupled with the gauge-covariance of the order parameter is enough to account for several of the striking phenomena of the superconducting state, independent of the choice of formalism: it doesn't matter whether $\mathcal{F}$ is a Lagrangian or a Free Energy. For the most part, this portion of the paper is a more elementary presentation of ideas originally developed in Refs.  \cite{weinbergbook, weinbergpaper}. Afterward, we will consider the central role of the Anderson-Higgs mechanism in the story of gauge symmetry breaking. 

Since the electromagnetic field, to which the superconductor is coupled, is intrinsically relativistic, it is convenient to adopt relativistic language. Throughout, we will use the  metric with signature $(+,-,-,-)$, which is the standard in particle physics, and sums over repeated indices are always implied. Spacetime coordinates are written as $x^\mu = (t, \v{x})$ where $\mu = 0,1,2,3$, derivatives with respect to $x^\mu$ are written as $\del_\mu$, and the scalar and vector potentials are combined into the four-potential $A_\mu = (\varphi, -\v{A})$. Throughout the remainder of this section we will make use of this relativistic notation, but at no point will we actually assume the theory we are discussing is Lorentz invariant: we are merely using this notation as a convenient shorthand in discussing a presumably non-relativistic system. The reader unfamiliar with electromagnetism from the perspective of classical field theory is encouraged to consult the literature. \cite{classical,fmunu,jackson}

In general, the effective action may depend on the order parameter field $\Delta$, its complex conjugate $\Delta^\star$, and their derivatives $\del_\mu \Delta$, $\del_\mu \Delta^\star$, as well as the electromagnetic field via dependences on the potential $A_\mu$ and its derivatives $\del_\mu A_\nu$. So, we expect to have some functional of these quantities, $\mathcal{F}[\Delta, \Delta^\star, \del_\mu \Delta, \del_\mu \Delta^\star, A_\mu, \del_\mu A_\nu]$. However, gauge invariance strongly constrains the form of this functional. Clearly the product $\Delta^\star \Delta$ is gauge invariant and we may use it to construct terms in the effective action, but the derivatives of the order parameter transform as
\begin{equation}
\del_\mu \Delta \mapsto \del_\mu \big( \e^{2ie\alpha}\Delta) = \e^{2ie\alpha}( \del_\mu + 2ie \del_\mu \alpha) \Delta,
\end{equation}
which is clearly not gauge invariant. To construct gauge invariant derivative terms we must introduce the \emph{gauge covariant derivative}, $D_\mu \equiv \del_\mu - 2ie A_\mu$. In our relativistic notation, a gauge transformation of the potentials is written $A_\mu \mapsto A_\mu + \del_\mu \alpha$, so the covariant derivative of the order parameter transforms as 
\begin{equation}
D_\mu \Delta = (\del_\mu - 2ieA_\mu)\Delta \mapsto \e^{2ie\alpha}\left(\del_\mu + 2ie\,\del_\mu \alpha - 2ie(A_\mu + \del_\mu \alpha) \right)\Delta = \e^{2ie\alpha}D_\mu \Delta.
\end{equation}
One can easily show that the complex conjugate $(D_\mu \Delta)^\star$ transforms as $(D_\mu \Delta)^* \mapsto \e^{-2ie\alpha}(D_\mu \Delta)^\star$, so we can form gauge invariant terms from the product $(D_\mu \Delta)^\star D^\mu \Delta$. Thus, gauge invariance constrains the effective action to be a functional of the form $\mathcal{F}[\Delta^\star \Delta, (D_\mu \Delta)^\star D^\mu \Delta]$. 

In fact, we can be slightly less restrictive by  re-parameterizing the two degrees of freedom of the complex order parameter as an amplitude and a phase, $\Delta = |\Delta | \e^{i\phi}$. Evidently, any functional of the amplitude will be gauge invariant. Expressed in terms of the amplitude and phase fields, the derivatives can be written
\begin{equation} \label{derivdelta}
\del_\mu \Delta = \del_\mu \big(|\Delta| \e^{i\phi}\big) = \left( i \del_\mu \phi + \frac{\del_\mu |\Delta|}{|\Delta|} \right) \Delta. 
\end{equation}
Recalling our discussion in section \ref{glsection}, the amplitude of the order parameter will be fixed at the non-zero value which minimizes the system's energy, and fluctuations around this minimum will be gapped and energetically costly. Since we are concerned with the low energy behavior, it is safe to assume that $\del_\mu |\Delta| \ll |\Delta|$, and we may drop the second term in Eq. (\ref{derivdelta}) so $\del_\mu \Delta \approx i \del_\mu \phi \Delta$. Note that under a gauge transformation $\Delta \mapsto \e^{2ie\alpha} \Delta$, and thus the phase transforms as $\phi \mapsto \phi + 2e\alpha$. 

In this representation the gauge covariant derivative can be written as $D_\mu \Delta = i(\del_\mu \phi - 2eA_\mu)\Delta$, and we may recognize the term in parentheses to be gauge invariant: $\del_\mu \phi - 2e A_\mu \mapsto \del_\mu (\phi + 2e\alpha)- 2e(A_\mu +\del_\mu \alpha) = \del_\mu \phi - 2e A_\mu$. So, the effective action can be a functional of only the amplitude $|\Delta|$ and the gauge invariant combination $\del_\mu \phi - 2e A_\mu$ such that we have $\mathcal{F}[|\Delta|, \del_\mu \phi - 2e A_\mu]$. In fact, there are no other gauge-invariant combinations of the fields that we could write down, and the previous functional form we considered is a special case of this structure. Thus, we have determined the most general constraints that gauge invariance imposes on the effective action. 

Now, given the effective action we may determine the ground state by the appropriate variational principle. By assumption, the ground state is attained for a non-zero value of $|\Delta|$, which as we saw in the previous section is the origin of ``broken'' gauge symmetry. As far as the dependence on $\del_\mu \phi - 2e A_\mu$, stability requires that the minimum of $\mathcal{F}$ occur for a finite value of its arguments, and we further expect it to be a quadratic function near the minimum, \cite{currentnote}
\begin{equation}
\mathcal{F} \sim K \big( \del_\mu \phi - 2e A_\mu\big)^2,
\end{equation}
for some prefactor $K$. Then, $\mathcal{F}$ will be minimized when
\begin{equation} \label{key}
\del_\mu \phi = 2e A_\mu.
\end{equation}
This condition on the ground state follows directly from the requirement that the effective action be gauge invariant, and it is sufficient to derive several of the properties of the superconducting state.

\subsection{Perfect Conductivity}
The timelike component of Eq. (\ref{key}) requires that inside a superconductor $\del_t \phi = 2e \varphi$, where $\varphi$ is the scalar potential. Then, for any time independent field configuration for which $\del_t \phi = 0$, we must have
\begin{equation}
\varphi = 0.
\end{equation}
So, the voltage between any two points inside the superconductor is $V = \varphi(\v{r}_2)-\varphi(\v{r}_1) = 0$. One time independent field configuration is a steady current $\v{j} = \text{const}$, which by the above argument must occur with zero potential difference. A finite current maintained at zero voltage implies the conductivity is infinite, since $\v{j} = \sigma V$ with $V = 0$ and $\v{j} \neq 0$ requires the conductivity $\sigma$ be infinite. Thus, superconductors may maintain persistent, dissipationless currents. \cite{weinbergpaper}

\subsection{Flux Quantization}
Consider a thick superconducting ring, and a closed contour $\gamma$ traversing it. The magnetic flux $\Phi$ through the surface $S$ bounded by the contour is
\begin{align}
\begin{split}
\Phi &= \int_S \v{B \cdot}\mathrm{d}\v{S}\\
&= 	\int_S (\nabla \v{\times A})\cdot \mathrm{d}\v{S} \\
&= \oint_\gamma \v{A \cdot} \mathrm{d}\v{s} \\
&= \frac{1}{2e} \oint_\gamma \nabla \phi \, \cdot \mathrm{d}\v{s} \\
&= \frac{1}{2e} \Delta \phi,
\end{split}
\end{align}
where we used Stoke's theorem to get from the second equality to the third, and Eq. (\ref{key}) to get from the third to the fourth. The final equality comes from the fact that the closed line integral over the gradient of $\phi$ is simply the difference between $\phi$ at the beginning and end of the path. Normally, such an integral must be zero, but since $\phi$ is an angle defined only modulo $2\pi$, we may have $\Delta\phi = 2\pi n$ for any integer $n$. Thus, the flux through the loop is quantized as 
\begin{equation}
\Phi = \frac{\hbar \pi n}{e}.
\end{equation}

\section{The Meissner Effect and the Anderson-Higgs Mechanism} \label{higgs}
One could use the spatial components of Eq. ($\ref{key}$) to show $\v{B} = \nabla \times \v{A} = \nabla \times \nabla \phi = 0$, which is to say that a magnetic field cannot exist in the ground state of a superconductor. However, while valid, this overlooks the rich physics underlying the Meissner effect, namely the Anderson-Higgs mechanism. In fact, this mechanism is the core of the gauge symmetry breaking story, as we will presently see.

Recall the simple Ginzburg-Landau theory from section \ref{glsection} where we found the ground state of the superconductor is characterized by a nonzero order parameter amplitude, $|\Delta| = \Delta_0$ and the absence of electromagnetic fields. The ground state configuration is then given by
\begin{equation} 
\Delta = \Delta_0 \e^{i \phi}\,, \qquad \v{E} = \v{B} = 0.
\end{equation}
The existence of an infinite set of ground states, each with a different phase $\phi$ led us to believe that gauge symmetry was spontaneously broken.

However, we must not forget that the order parameter does not completely specify the state of the system, we also have the electromagnetic field. Having seen that the potentials $A_\mu$ are directly tied to the order parameter, and are in some sense the fundamental entity, we will use $A_\mu$ to specify the electromagnetic field configuration. We can recover $\v{E}$ and $\v{B}$ from the components of the field strength tensor, \cite{fmunu} $F_{\mu \nu} = \del_\mu A_\nu - \del_\nu A_\mu$, as $E_i = F_{0i}$ and $B_i = \frac{1}{2}\epsi_{ijk}F_{jk}$, where the roman indices run over the spatial components $1,2,3$. In the ground state, we have $\v{E} = \v{B} = 0$ when $F_{\mu \nu} = 0$, which is achieved when the potential is a ``pure gauge,'' $A_\mu = \del_\mu \beta$, for any scalar field $\beta$. This is simply to say that the potential is related to $A_\mu = 0$ by a gauge transformation, and thus represents the same physical state. The important point is that any choice of $\beta$ is valid, and thus there are an infinite number of potentials which describe the same physical state of $\v{E} = \v{B} = 0$. The ground state of the superconducting system is then given by
\begin{equation} \label{gsmfd}
\Delta = \Delta_0 \e^{i \phi}\, , \qquad A_\mu = \del_\mu \beta.
\end{equation}
Now, suppose we perform a gauge transformation on this state, under which the potential transforms $A_\mu \mapsto A_\mu + \del_\mu \alpha$, and the phase of the order parameter transforms as $\phi \mapsto \phi + 2e\,\alpha$. The ground state in Eq. (\ref{gsmfd}) then becomes
\begin{equation}
\Delta = \Delta_0 \e^{i(\phi + 2e\alpha)}, \qquad A_\mu = \del_\mu (\alpha + \beta).
\end{equation}
Evidently, by performing a gauge transformation we may rotate the phase of $\Delta$. Further, since our choice of $\alpha$ is arbitrary, we may rotate the phase into any value we wish. Specifically, we can perform a gauge transformation to move from one ground state of fixed phase to another. We also know that two field configurations related by a gauge transformation represent the same physical state, and thus conclude that the all of the seemingly degenerate ground states in Eq. (\ref{gsmfd}) are merely \emph{different descriptions of the same physical state}. This is the essence of ``broken'' gauge symmetry: there is only one physical ground state, wherein $|\Delta| = \Delta_0$ and $\v{E} = \v{B} = 0$, but an infinite number of gauge equivalent ways to describe it. The existence of multiple order parameter configurations to describe the same ground state \emph{looks} extremely similar to the ground state degeneracy arising from a spontaneously broken symmetry, which has led to the misleading terminology of ``spontaneously broken gauge symmetry.'' Crucially, one must note that at no point did gauge symmetry ever actually ``break:'' one can verify that every step of our analysis (and all of our discussion to follow) is perfectly gauge invariant, as any consistent physical theory must be. In fact, not only is the violation of gauge symmetry nonsensical, but, at least in lattice gauge theories, it is not possible for such a symmetry to be spontaneously broken due to the famous Elitzur theorem. \cite{elitzur}

Inseparable from this illusion of gauge symmetry breaking is the Anderson-Higgs mechanism, by which the spectrum of the gauge field acquires a gap. In the particle physics vernacular this corresponds to the gauge field acquiring a mass, and as this language has bled into the condensed matter literature, it is often said that the ``photon acquires a mass.'' 

To illustrate the Anderson-Higgs mechanism in its most natural context, let us consider the Lorentz invariant analogue of a superconductor: the Abelian-Higgs model, defined by the Lagrangian \cite{fmunu}
\begin{equation}
\lag = (D^\mu \Delta)^* D_\mu \Delta - \frac{1}{4}F_{\mu \nu}F^{\mu \nu} - V(|\Delta|).
\end{equation}
Although this model does not actually describe a superconductor, nor does it play any role in the Standard Model, it is an incredibly useful pedagogical tool to understand the Anderson-Higgs mechanism, especially as it applies to superconductors. We assume the potential $V$ is minimized by a nonzero value of the order parameter amplitude, giving rise to the gauge-equivalent ground states in Eq. (\ref{gsmfd}) just as in the non-relativistic case. By parameterizing the order parameter by its amplitude and phase, the covariant derivative can be written $D_\mu \Delta=(\del_\mu \phi - 2eA_\mu) \Delta$, and thus the Lagrangian becomes
\begin{equation}
\lag = |\Delta|^2 \big( \del_\mu \phi - 2eA_\mu \big)^2   - \frac{1}{4}F_{\mu \nu}F^{\mu \nu} - V,
\end{equation}
which we may again rewrite by defining a new vector field,
\begin{equation}
\tilde{A}_\mu \equiv A_\mu - \frac{1}{2e}\del_\mu \phi.
\end{equation}
Note that this is \emph{not} a gauge transformation: $\tilde{A}_\mu$ is a new gauge invariant vector field, since under a gauge transformation
\begin{equation}
\tilde{A}_\mu \mapsto A_\mu + \del_\mu \alpha - \frac{1}{2e}\del_\mu \big(\phi + 2e \, \alpha \big) = A_\mu - \frac{1}{2e}\del_\mu \phi = \tilde{A}_\mu.
\end{equation}
Noting that $F_{\mu \nu}$ is invariant under this redefinition, the Lagrangian becomes
\begin{equation}
\lag = (2e)^2 |\Delta|^2 \, \tilde{A}_\mu \tilde{A}^\mu   - \frac{1}{4}F_{\mu \nu}F^{\mu \nu} - V .
\end{equation}
In the field theory literature, the term quadratic in $\tilde{A}_\mu$ implies this vector field has a mass of $m^2 = 2e^2 |\Delta|^2$. To make this explicit, we begin by finding the Euler-Lagrange equations of motion for the system,
\begin{align}
\deriv{\lag}{\tilde{A}^\nu} - \del_\mu \deriv{\lag}{(\del_\mu \tilde{A}^\nu)} &= 0 \\
m^2 \tilde{A}_\nu + \del^\mu F_{\mu \nu} &= 0 \\
m^2 \tilde{A}_\nu + \del^\mu \big(\del_\mu \tilde{A}_\nu - \del_\nu \tilde{A}_\mu \big) &= 0 \\
m^2 \tilde{A}_\nu + \del^\mu \del_\mu \tilde{A}_\nu - \del_\nu \del^\mu \tilde{A}_\mu &=0 .\label{eom1}
\end{align}
To simplify this expression, we note the current is given by \cite{classical}
\begin{equation} \label{currenta}
j_\mu = - \deriv{\lag}{\tilde{A}^\mu} = -m^2 \tilde{A}_\mu. 
\end{equation}
For this current to be conserved, we must have $\del^\mu j_\mu = 0$, and thus $\del^\mu \tilde{A}_\mu = 0$, which requires the last term in Eq. (\ref{eom1}) to vanish. The equation of motion is then simply
\begin{equation} \label{kg}
\big(\del_\mu \del^\mu + m^2 \big)\tilde{A}_\mu = 0.
\end{equation}
This is the Klein-Gordon equation, which describes a classical field of mass $m$. This is made clear by Fourier transforming the field,
\begin{equation}
\tilde{A}_\mu (x) = \int \df^4 x \; \e^{i p_\mu p^\mu}\, \tilde{A}_\mu(p),
\end{equation}
where $p^\mu = (E, \v{p})$. Then, the Klein-Gordon equation becomes
\begin{align}
(-p_\mu p^\mu + m^2)\tilde{A}_\mu (p) &= 0 \\
-p_\mu p^\mu + m^2 &= 0 \\
E^2 = p^2 &+ m^2,
\end{align}
which we recognize as the relativistic dispersion of a particle with mass $m$. If we restrict ourselves to considering only the spacelike components of Eq. (\ref{kg}) (due to the additional structure of Lorentz invariance, the timelike component will be substantially different from the nonrelativsitic case, to which we will momentarily compare our results), and further specialize to time-independent scenarios, we have
\begin{equation}
\big( -\nabla^2 + m^2\big)\v{\tilde{A}} = 0.
\end{equation}
Taking the curl of this equation and noting that $\nabla \times \v{\tilde{A}} = \v{B}$ since $\nabla \times (\nabla \phi) =0$, we find an equation for the magnetic field,
\begin{equation}
\big(-\nabla^2 + m^2 \big)\v{B} = 0,
\end{equation}
which notably does not admit a constant solution, implying a uniform magnetic field cannot exist inside the superconductor. In fact, at the interface between a superconductor and the vacuum, we expect the field to decay exponentially like $\sim \e^{-mr}$, which is strongly reminiscent of the Meissner effect.

To be more physically accurate, let us return to superconductivity from the perspective of our original, non-relativistic Ginzburg-Landau free energy from Eq. (\ref{glexp}), which we may rewrite as
\begin{equation}
F = \int \df^3 x \; \frac{|\Delta|^2}{2m^\star} \big(\nabla \phi - 2e \v{A} \big)^2 + \frac{1}{2}\v{B}^2 + V(|\Delta|),
\end{equation}
or, in terms of the vector field $\v{\tilde{A}}$,
\begin{equation}
F = \int \df^3 x \; \frac{(2e)^2 |\Delta|^2}{2m^\star} \v{\tilde{A}}^2 + \frac{1}{2}\v{B}^2 + V(|\Delta|).
\end{equation}
Similarly to the relativistic case, the current is given by \cite{colemangl}
\begin{equation}
\v{j} = -\fderiv{F}{\v{\tilde{A}}} = - \frac{(2e)^2 |\Delta|^2}{m^\star} \v{\tilde{A}},
\end{equation}
and by Maxwell's equation $\nabla \times \v{B} = \v{j}$, we have
\begin{equation}
\nabla \times \v{B} = - \frac{(2e)^2 |\Delta|^2}{m^\star} \v{\tilde{A}}.
\end{equation}
Taking the curl of both sides, and using $\nabla \cdot \v{B} = 0$, we find
\begin{equation}
-\nabla^2 \v{B} = -\frac{(2e)^2 |\Delta|^2}{m^\star} \v{B},
\end{equation}
from which we again see that a uniform magnetic field is not admitted inside a superconductor. If we consider a superconductor occupying the half space $x>0$ and vacuum in $x<0$, the magnetic field will be given by
\begin{equation}
\v{B} = \v{B}_0 \e^{-x/\lambda}, \qquad \lambda^{-2} = \frac{(2e)^2 |\Delta|^2}{m^\star},
\end{equation}
where the parameter $\lambda$ is the \emph{London Penetration Depth}, which in SI units is
\begin{equation}
\lambda = \sqrt{\frac{m_e}{2\mu_0 e^2|\Delta|^2 } }.
\end{equation}
Notice that this is completely analogous to relativistic theory, with the role of the Anderson-Higgs generated mass being played by the inverse penetration depth. The Anderson-Higgs mechanism sketched here was initially put forth in the context of superconductivity by Anderson, \cite{andersonhiggs} but has risen to fame in particle physics, wherein this same mechanism generates the masses of the W and Z bosons. \cite{massnote} However, due to the non-Abelian nature of the ``broken'' gauge group in the electroweak theory, some aspects of the dynamics differ from the case we have discussed here. Further, just as the Higgs boson was recently discovered at the Large Hadron Collider, there have also been experimental observations of Higgs modes in superconductors. \cite{higgsexp}

\section{Conclusion}
Within the mean field approximation, the formation of Cooper pairs enables gauge covariant operators such as the order parameter $\Delta \sim \langle \Psi \Psi \rangle$ to acquire a nonzero expectation value in the ground state of the system. The order parameter amplitude being nonzero in the ground state allows the phase $\phi$ to be well-defined but arbitrary, which in turn leads to the apparent existence of a degenerate ground state manifold. However, due to the coupling with the electromagnetic field, all of these ground state solutions are in fact gauge equivalent descriptions of the same physical state. The excitations of the gauge field away from this ground state are gapped due to the Anderson-Higgs mechanism, which is responsible for the Meissner effect. In fact, we have seen that the gauge principal is sufficient to account for many of the striking phenomena observed in superconductors. In all, the question of whether or not gauge symmetry ``breaks'' in a superconductor is one of linguistics, but it is unambiguous that gauge invariance is absolutely central to the physics of superconductors.

\section*{Acknowledgements}

The author would like to thank Tom Cohen, Chris Lobb,  Rick Greene, and Sankar Das Sarma for productive conversations instrumental to clarifying his thinking and providing feedback on the manuscript. This work is partially supported by the NSF award DMR-1708334 and the Maryland Center for Nanophysics and Advanced Materials (CNAM).


\begin{thebibliography}{99}
\bibitem{lagnote} This is because the Lagrangian has a term $\sim q\v{A}\cdot\dot{\v{x}}$ coupling the particle to the field. The canonical momentum $\v{p} = \del L/\del \dot{\v{x}}$ then has an additional term $\sim q\v{A}$ which must be subtracted out to get the ``physical'' momentum $m\dot{\v{x}}$.
\bibitem{derivnote} Since $\alpha(\v{x},t)$ is dependent on space and time, derivatives acting on the gauge transformed wave-functions give $\nabla (\e^{iq\alpha}\psi) = \e^{iq\alpha}\nabla \psi + iq (\nabla \alpha)\e^{iq\alpha} \psi$, and similarly for time derivatives.
\bibitem{Neignote} Since for any system of definite particle number the energy eigenstates will be simultaneous eigenstates of the number operator, there is no way the system can time evolve into a state with different particle number. Put another way, since the number operator commutes with the Hamiltonian, particle number must be conserved.
\bibitem{eqpnote} Even though the pairs in the condensate are comprised of electrons, we treat them as different objects. To be more precise, the electrons we speak of are quasiparticles, not bare electrons.
\bibitem{saddleptnote} Expanding the free energy, which is known to be discontinuous at a phase transition, as an analytic series near the discontinuity can only be justified if we understand the free energy as the result of a saddle point approximation of the appropriate partition function. See Ref  \cite{kardar}.
\bibitem{rnote} Details of the Ginzburg-Landau theory, which we have glossed over, ensure that in the superconducting state $r < 0$, so the gap magnitude $\Delta^\dag \Delta = -r/u$ is a positive quantity. See Ref.  \cite{colemangl}
\bibitem{currentnote} Once could conceivably ask why the quadratic function must be centered about zero, i.e. why could we not have something like $\mathcal{F} \sim K[(\del_\mu \phi - 2eA_\mu)^2-j^2]^2$? The effective action would then be minimized when $j_\mu = \del_\mu \phi - 2e A_\mu$, which we know from Eq. (\ref{currenta}) represents current flow.
\bibitem{massnote} The Anderson-Higgs mechanism also generates the mass of quarks and electrons due to a Yukawa coupling that has no analogue in superconductors, and thus is not discussed here.
\end{thebibliography}
\end{document}